\documentclass[sigconf]{acmart}
\immediate\write18{texcount -inc -sub=section 00_main.tex > wordcount.txt}
\AtBeginDocument{%
  }

\copyrightyear{2026}
\acmYear{2026}
\setcopyright{cc}
\setcctype{by}
\acmConference[CHI '26]{Proceedings of the 2026 CHI Conference on Human Factors in Computing Systems}{April 13--17, 2026}{Barcelona, Spain}
\acmBooktitle{Proceedings of the 2026 CHI Conference on Human Factors in Computing Systems (CHI '26), April 13--17, 2026, Barcelona, Spain}
\acmPrice{}
\acmDOI{10.1145/3772318.3791513}
\acmISBN{979-8-4007-2278-3/2026/04}

\usepackage{tabularx,multirow,array,makecell}
\usepackage{varwidth}

\usepackage{verbatim}

\newcommand{\ai}{\textcolor{brown!80}{\faRobot}}
\newcommand{\nw}{\textcolor{darkgray!80}{\faGlobeAsia}}

\newcommand{\aioff}{\textcolor{gray!30}{\faRobot}}
\newcommand{\nwoff}{\textcolor{gray!30}{\faGlobeAsia}}

\newcommand{\added}[1]{#1}

\usepackage{xspace}
\newcommand{\parabold}[1]{%
  \par\vspace{1ex}
  \noindent\textbf{#1. }%
}

\newcounter{recommendationctr}
\newcommand{\recommendation}[2]{%
  \refstepcounter{recommendationctr}%
  \par\vspace{0.75ex}%
  \noindent\textbf{G\therecommendationctr:} \textbf{\textit{#1}}#2%
  \vspace{0.75ex}%
}

\newcommand{\vetbot}{VetBot\xspace}
\newcommand{\farmerchat}{FarmerChat\xspace}
\newcommand{\prompts}{PROMPTS\xspace}
\newcommand{\cataractbot}{CataractBot\xspace}
\newcommand{\ashabot}{ASHABot\xspace}
\newcommand{\padhai}{ReadAI\xspace}
\newcommand{\shiksha}{Shiksha Copilot\xspace}
\newcommand{\suvas}{LegalTranslateAI\xspace}

\newcommand{\promptsaidev}{P11\xspace}
\newcommand{\farmerchataidev}{P16\xspace}
\newcommand{\farmerchatdomain}{P14\xspace}

\newcommand{\padhaidomain}{P6\xspace}
\newcommand{\shikshadomain}{P8\xspace}
\newcommand{\shikshadomaintwo}{P5\xspace}
\newcommand{\shikshaaidev}{P2\xspace}
\newcommand{\suvasdomain}{P9\xspace}
\newcommand{\suvasaidev}{P17\xspace}
\newcommand{\ashabotdomain}{P12\xspace}
\newcommand{\aiexpertone}{P4\xspace}

\usepackage{fontawesome5}
\usepackage{xcolor}
\usepackage{colortbl}

\newcommand{\tagbox}[2]{%
  \begingroup
  \setlength{\fboxsep}{2pt}
  \colorbox{#1!20}{%
    \textcolor{#1!60!black}{%
      \strut\hspace{0.3em}#2\hspace{0.3em}%
    }%
  }%
  \endgroup
}

\newcommand{\inweight}[1]{%
  \tagbox{purple}{\faProjectDiagram\ #1}%
}

\newcommand{\incontext}[1]{%
  \tagbox{blue}{\faAlignLeft\ #1}%
}

\begin{document}

\title{Designing Culturally Aligned AI Systems For Social Good in Non-Western Contexts}

\title{Designing Culturally Aligned AI Systems For Social Good in Non-Western Contexts}

\author{Deepak Varuvel Dennison}
\orcid{0009-0004-8355-5024}
\affiliation{%
  \institution{Cornell University}
  \city{Ithaca}
  \state{New York}
  \country{USA}
}
\email{dv292@cornell.edu}

\author{Mohit Jain}
\affiliation{%
  \institution{Microsoft Research}
  \city{Bangalore}
  \country{India}}
\email{mohja@microsoft.com}

\author{Tanuja Ganu}
\affiliation{%
  \institution{Microsoft Research}
  \city{Bangalore}
  \country{India}}
\email{tanuja.ganu@microsoft.com}

\author{Aditya Vashistha}
\affiliation{%
  \institution{Cornell University}
  \city{Ithaca}
  \state{New York}
  \country{USA}
}
\email{adityav@cornell.edu}

\renewcommand{\shortauthors}{Varuvel Dennison et al.}

\begin{abstract}
AI technologies are increasingly deployed in high-stakes domains such as education, healthcare, law, and agriculture to address complex challenges in non-Western contexts. This paper examines eight real-world deployments spanning seven countries and 18 languages, combining 17 interviews with AI developers and domain experts with secondary research. Our findings identify six cross-cutting factors — Language, Institution, Safety, Task, End-User Demography, and Domain — that structured how systems were designed and deployed. These factors were shaped by Sociocultural (diversity, practices), Institutional (resources, policies), and Technological (capabilities, limits) influences. \added{We find that building effective AI systems required extensive collaboration between AI developers and domain experts, with human resources proving more critical to achieving safe and effective outcomes in high-stakes domains than technological expertise alone. Additionally, we present 12 guidelines synthesizing these dynamics for designing AI for social good systems that are culturally grounded, equitable, and responsive to the needs of non-Western contexts.}
\end{abstract}



\maketitle


\section{Introduction}\label{sec:introduction}

Artificial intelligence (AI) technologies are spreading quickly across the globe with their largest user growth in non-Western contexts \cite{liang_widespread_2025, loewenglobal}. AI applications now extend beyond productivity or creative tools into education \cite{choi_are_2024, dennison_teacher-ai_2025}, healthcare \cite{mateen_trials_2025, ramjee_cataractbot_2025, ramjee_ashabot_2025}, agriculture \cite{singh_farmerchat_2024, shepherd_virtual_nodate}, and law \cite{becher_lexoptima_2025}. In non-Western contexts, this shift carries particular weight as these systems address high-stakes, wicked problems where model outputs can directly affect people’s health, learning, and livelihoods.

Recent research shows that the integration of AI into high-stakes environments have promise. AI can augment information processing \cite{fui-hoon_nah_generative_2023, moulaei_generative_2024}, strengthen education and training \cite{baidoo-anu_education_2023}, boost productivity \cite{al_naqbi_enhancing_2024}, and open new creative possibilities \cite{zhou_generative_2024}. Yet the risks are profound. Prior work shows that they reproduce gender, racial, caste, and ableist biases \cite{aneja_beyond_2025, gallegos_bias_2024, seth_how_2018, phutane_cold_2025}, lack cultural awareness \cite{agarwal_ai_2025, chiu_culturalteaming_2024, tao_cultural_2024}, and provide outputs that are more accurate and relevant for Western users than for those in non-Western contexts \cite{bhagat_richer_2024, mirza_global-liar_2024}. Beyond these representational harms, they risk entrenching colonial epistemologies by sidelining diverse knowledge systems \cite{dewitt_prat_decolonizing_2024}.

Successfully using AI for social good in non-Western contexts thus requires more than deploying mainstream models; it demands deliberate cultural and linguistic alignment and developing sociotechnical infrastructures that make these systems culturally aligned, contextually relevant, and safe to use. However, given the rapid integration of AI into these settings, and despite the urgency and risks that this integration can pose, little empirical work has examined how AI systems are actually designed and implemented in such settings. We lack a grounded understanding of the tradeoffs developers face, the forms of expertise required, and the human labor that underpins this adaptation. In this paper, we ask three questions:

\begin{itemize}
    \item[\textbf{RQ1:}] How are AI systems adapted and contextualized for high-stakes domains in non-Western contexts?

      \item[\textbf{RQ2:}] What sociocultural, institutional, and technological influences shape design and deployment?

    \item[\textbf{RQ3:}] What lessons can guide the design and implementation of culturally aligned, socially beneficial AI systems in these settings?
\end{itemize}

\added{
To answer these questions, we conducted semi-structured interviews with AI developers and domain experts responsible for systems deployed at scale, reaching from hundreds to millions of users across education, healthcare, agriculture, and law. Our analysis identifies six cross-cutting factors that shaped how culturally aligned AI systems were built and deployed: \textsc{\textbf{L}anguage}, \textsc{\textbf{I}nstitution}, \textsc{\textbf{S}afety}, \textsc{\textbf{T}ask}, \textsc{\textbf{E}nd-User Demography}, and \textsc{\textbf{D}omain} (\textbf{LISTED}). Language required grounding models in local linguistic realities through technical adaptations and community-built data. Institutions determined whether systems could move beyond pilots, since institutional authority and alignment shaped adoption at scale. Safety depended on human oversight and technical safeguards, mediating trust. Task demands introduced practical constraints such as latency, modality, and environmental conditions, often requiring customized AI workflows. End-User Demography shaped system relevance through variation in literacy, age, gender and cultural norms. Domain anchored reliability by defining what counted as correct or trustworthy in each field, which required sustained involvement from domain experts. These factors intersected and reinforced each other, surfacing persistent tradeoffs.


We synthesize these findings into three higher-level influences that shaped the LISTED factors: \textbf{\textsc{Sociocultural}, \textsc{Institutional}, and \textsc{Technological}}. Together, the LISTED factors and the higher-level influences explain which systems are possible to build, which communities they reach, and how they endure over time in non-Western contexts. Across all cases, one consistent finding emerged: human labor was central to making the systems work effectively. Developers and field workers were responsible for the curation, adaptation, and oversight that made these systems usable, contextual, and safe. This paper makes three contributions:
\begin{itemize}
\item An empirical account of how AI-driven applications are designed and implemented in high-stakes, non-Western domains.
\item A framework detailing six factors that shape AI system design and three overarching influences that govern how these factors manifest in practice. Together, these components clarify why certain design choices emerge, how systems become usable for diverse populations, and what enables long-term adoption and legitimacy.
\item Twelve practical guidelines derived from these findings to support practitioners in building culturally aligned AI systems for high-stakes, non-Western settings.
\end{itemize}
}


\section{Related Work}\label{sec:related_work}
HCI and ICTD scholarship has long emphasized that technologies are never value-neutral but are deeply embedded in cultural, political, and economic contexts~\cite{heimgartner_culturally-aware_2018, hayes_inclusive_2020, pereira_value-oriented_2015, salgado_cultural_2015}. Early ICTD work warned that interventions fail when detached from local realities, noting that technology tends to amplify existing intent and capacity rather than substitute for them~\cite{toyama_geek_2015}. The One Laptop per Child (OLPC) project illustrates this point: despite its visionary appeal, OLPC laptops were often underused or repurposed for games because teachers were untrained, curricula were not adapted, and basic infrastructure like electricity and internet access was missing~\cite{ames_charisma_2019}. As Ames argues in The Charisma Machine, techno-solutionist approaches risk overlooking structural inequities and local context, leading to disappointing outcomes. Postcolonial computing extended this critique, showing how design practices frequently reinscribe Western epistemologies and marginalize alternative ways of knowing~\cite{irani_postcolonial_2010, philip_postcolonial_2012}. Together, these insights call for careful cultural adaptation — technologies must be grounded in local needs and designed with attention to the sociocultural, socioeconomic, and sociopolitical forces that shape their adoption and use. This is especially crucial for technologies meant to address complex social problems, where failures can have outsized and harmful effects on already marginalized communities~\cite{wyche_using_2020, pyram_future_2024}.

With the rise of LLMs, HCI and ICTD scholars have increasingly turned to these technologies to tackle complex social problems in high-stakes domains such as education, healthcare, law, and agriculture — an area often framed as AI for Social Good (AI4SG) \cite{cowls_designing_2019, floridi_how_2020, shi_artificial_2020}. A growing body of work shows that LLMs have potential to improve teaching and training~\cite{peters_beyond_2025}, address health information inequities~\cite{fuentes_opportunities_2024}, and boost productivity~\cite{mutambara_artificial_2025}.  But these promises come with significant risks. Research also shows that LLMs routinely reproduce identity-based biases~\cite{aneja_beyond_2025, gallegos_bias_2024, seth_how_2018, phutane_cold_2025},  lack cultural awareness~\cite{agarwal_ai_2025, chiu_culturalteaming_2024, tao_cultural_2024}, and can produce outputs that are systematically less accurate and relevant for users in non-Western contexts—creating quality-of-service harms~\cite{bhagat_richer_2024, mirza_global-liar_2024}. \citet{anuyah_cultural_2024} review cultural considerations for AI design in the Global South and show that most models still reflect Western-centric assumptions~\cite{prabhakaran_cultural_2022, zhou_bias_2024}. Similarly, \citet{karamolegkou_nlp_2025} identify persistent challenges in NLP for social good, including data scarcity and bias in low-resource languages, misaligned evaluation metrics, safety and ethical risks, limited cross-cultural generalization, and infrastructural gaps. Beyond these representational harms, research shows that LLMs risk entrenching colonial epistemologies by sidelining diverse knowledge systems \cite{dewitt_prat_decolonizing_2024}. These limitations make cultural and contextual alignment essential: without it, AI systems risk not only technical failure but also the erosion of trust and legitimacy in the very communities they aim to serve.

In response to such risks and harms, AI4SG scholars have 
argued for culturally responsive, community-driven AI solutions that prioritize equity and inclusivity~\cite{wasi_ai_2025,bondi_envisioning_2021, delgado_participatory_2023, corbett_power_2023, stawarz_co-designing_2023}. Yet, despite these calls, little is known about \textit{how} AI developers and domain experts navigate the complex sociotechnical, institutional, and cultural factors that shape these systems in practice and \textit{what} sociocultural, institutional, and technological factors shape their design and implementation. 

Our study addresses this gap by analyzing real-world AI deployments at scale across multiple domains — education, healthcare, agriculture, and law — in non-Western contexts. We identify six cross-cutting factors (Language, Institution, Safety, Task, End-User Demography, and Domain) that structure system design and implementation and synthesize them into three higher-order influences that explain adoption, legitimacy, and sustainability of these interventions. In doing so, we provide one of the first empirically grounded accounts of how AI for Social Good systems are built and maintained in high-stakes, non-Western contexts, offering concrete design recommendations for building culturally aligned, socially beneficial AI.

\section{Methods}\label{sec:methods}
To investigate how AI systems are built and deployed for non-Western populations in high-stakes domains, we examined real-world deployments using key-informant interviews and \added{secondary research drawing on existing reports and documentation across all of the projects studied}. The study protocol was approved by the Institutional Review Board (IRB), and all procedures adhered to established ethical research guidelines.

\subsection{AI Project Selection}
We used a purposive sampling to select AI for social good projects. The projects were selected based on the following criteria: (i) targeted a problem in a high-stakes domain, (ii) were deployed on the ground with active users in non-Western contexts, \added{(iii) were made available to users at no financial cost and operated by non-profit institutions, and} (iv) incorporated at least one AI technology as a core component. \added{In our study, we categorize mainstream language technologies as AI technologies. We focus specifically on Machine Translation (MT), Language Models, and Automatic Speech Recognition (ASR), as they constitute the most widely deployed forms of AI today and therefore provide a useful lens for understanding patterns that extend to other AI domains.} We also sampled across domains, geographies, languages, and institutional contexts to ensure diversity.

Table~\ref{tab:projects} shows the \textbf{eight projects} we selected spanning \textbf{four domains} (education, healthcare, agriculture, legal). \added{With the exception of two projects (\prompts and \farmerchat), all the projects were cross organizational collaborations, where a non-governmental organization (NGO) handled the implementation and a technology company handled the technical aspects. Aforementioned two organizations, however, had sufficient in-house capacity to develop AI technologies given the large scale of their operations.}

The projects varied widely in scope -- from pilot scale ($<\!500$ users) to large deployments (up to $\sim$350,000 users). The projects were based in \textbf{seven countries} (India, Bangladesh, Colombia, Kenya, Ethiopia, Nigeria, Ghana) and supported \textbf{18 languages}. Projects leveraged multiple delivery mediums (Web apps, Mobile apps, WhatsApp, SMS) for delivering the solution, and a range of AI technologies (LLMs; speech technologies for Text-to-Speech (TTS)/Speech-to-Text (STT); Neural Machine Translators (NMTs); transformer-based classifiers). Two of the projects -- \suvas and \padhai -- are pseudonymized in the paper to ensure anonymity as requested by the participants.


\begin{table*}[t]
\added{
\centering

\renewcommand{\arraystretch}{1.5}
\setlength{\tabcolsep}{4pt}
\begin{tabularx}{\textwidth}{%
  >{\raggedright\arraybackslash}p{2.35cm} 
  >{\raggedright\arraybackslash}p{1.5cm} 
  >{\raggedright\arraybackslash}p{1.5cm} 
  >{\raggedright\arraybackslash}p{2.25cm}   
  >{\raggedright\arraybackslash}X   
  >{\raggedright\arraybackslash}p{1.85cm}   
  >{\raggedright\arraybackslash}p{1.0cm}        
}

\toprule
\textbf{Project} & \textbf{Domain} & \textbf{AI Tech} & \textbf{Usage} & \textbf{Languages}  & \textbf{Geography} & \textbf{Year of Launch} \\
\midrule
\arrayrulecolor{gray!30}

\vetbot & Agriculture & LLM, STT, TTS & $<\!500$ users & Tamil, English, Kannada, Hindi, Malayalam & India & 2025 \\
\hline

\farmerchat & Agriculture & LLM, STT, TTS & $\sim$15{,}000 farmers & Swahili, Amharic, Hausa, Hindi, Odiya, Telugu, Kannada, English & Ethiopia, India, Kenya, Nigeria &  2023\\
\hline

\prompts & Healthcare & LLM, BERT Classifier & $\sim$350{,}000 women; $\sim$1.1M SMS/mo & Swahili, Hausa, Yoruba, Xhosa, Zulu & Ghana, Kenya, Nigeria & 2023 \\
\hline

\cataractbot & Healthcare & LLM, STT, TTS & $\sim$2000 patients & English, Kannada, Hindi, Telugu, Tamil, Urdu & India & 2024 \\
\hline

\ashabot & Healthcare & LLM, STT, TTS & $\sim$5{,}000 health workers & Hindi, English, Telugu, Marathi & India & 2024 \\
\hline

ReadAI* & Education & STT & $\sim$5,000 children & Hindi, Marathi & Colombia, India & 2025 \\
\hline

\shiksha & Education & LLM & $\sim$1{,}100 teachers & Kannada, Telugu, Hindi, English & India & 2024 \\
\hline

LegalTranslateAI* & Legal & NMT & $\sim$120,000 Court Judgments & Hindi, Kannada, Tamil, Telugu, Punjabi, Marathi, Gujarati, Malayalam, Bengali, Urdu & Bangladesh, India & 2019 \\

\bottomrule
\end{tabularx}
\begin{minipage}{\linewidth}
\raggedleft
*Project name pseudonymized
\end{minipage}

\caption{Overview of included projects}
\label{tab:projects}
}
\end{table*}


\subsection{Semi-structured Interviews}
\added{After shortlisting the projects, using a combination of cold emails and our personal network,  we contacted AI developers and domain experts working on these projects 
to serve as our primary informants. These two groups were selected because their complementary expertise in technical design and on-the-field implementation provide a comprehensive understanding of the challenges these projects experience when deployed on-the-ground. }

\added{\parabold{AI developers} We define AI developers as technologists directly responsible for building the AI components of the tool. Their work included integrating existing models and pipelines and, when necessary, fine-tuning or training new models. 
Often in these projects, small development teams took on multiple responsibilities, including both system engineering and AI development.}

\added{\parabold{Domain experts} We define domain experts as practitioners who worked closely with field staff, end users, and AI developers to ensure that the tool meet on-the-ground needs. Their responsibilities included communicating user requirements to the technical teams, coordinating deployment logistics, and monitoring how the system operated in local contexts and impacted on-the-ground workflows.}

We conducted interviews \added{between June and July 2025} with \textbf{17 participants} \added{(10 AI Developers and 7 Domain Experts) across different projects}. Interviews were semi-structured, lasting between 60 and 90 minutes, and conducted either via Zoom or in person (see Appendix \ref{appendix:ai-dev-interview-protocol} \& \ref{appendix:domain-expert-interview-protocol} for interview protocols). All participants provided informed consent and were assured that they could withdraw at any point. With explicit permission, interview sessions were audio-recorded and subsequently transcribed using the Whisper STT model, followed by manual cleaning to ensure accuracy.

\begin{table*}[t]
\centering
\setlength{\tabcolsep}{10pt}
\begin{tabularx}{0.65\textwidth}{l l l l l}
\toprule
\textbf{Participant ID} & \textbf{Domain} & \textbf{Role} & \textbf{Gender} & \textbf{Experience} \\
\midrule
P1   & Agriculture        & Domain Expert & Male   & 35 \\
P2   & Education          & AI Developer  & Male   & 3  \\
P3   & Agriculture        & AI Developer  & Female & 22 \\
P4   & Healthcare         & AI Developer  & Female & 15 \\
P5   & Education          & Domain Expert & Male   & 32 \\
P6   & Education          & AI Developer  & Female & 6  \\
P7  & Healthcare         & AI Developer  & Male   & 14 \\
P8   & Education          & Domain Expert & Female & 18 \\
P9   & Legal              & Domain Expert & Male   & 28 \\
P10  & Agriculture        & AI Developer  & Male   & 13 \\
P11  & Healthcare         & AI Developer  & Male   & 19 \\
P12  & Healthcare         & Domain Expert & Male   & 10 \\
P13  & Healthcare         & Domain Expert & Male   & 20 \\
P14  & Agriculture        & Domain Expert & Female & 18 \\
P15  & Healthcare         & AI Developer  & Male   & 2  \\
P16  & Agriculture        & AI Developer  & Male   & 18 \\
P17  & Legal              & AI Developer  & Male   & 14 \\
\bottomrule
\end{tabularx}
\caption{Participant Information}
\end{table*}

\subsection{Secondary Research}
We complemented interviews with secondary research \added{\cite{stewart1993secondary}} conducted beforehand, \added{drawing on published research papers, web articles, and AI model cards related to the chosen projects.} This provided background context, guided our interviews, and enabled us to triangulate information across multiple perspectives. To ensure accuracy, we confirmed and cross-checked these details with the key informants during interviews and through follow-up member-checking (see ~\S\ref{sec:analysis}).

\subsection{Analysis} \label{sec:analysis}
We conducted a qualitative thematic analysis \cite{fereday_demonstrating_2006} of all study materials. The research team iteratively developed a coding system that combined inductive themes emerging from the data with deductive categories grounded in an AI-adaptation taxonomy, following \citet{liu_culturally_2025}, which distinguishes \emph{in-weight} and \emph{in-context} adaptations for enabling cultural alignment in AI systems.\footnote{\textbf{In-weight} adaptations modify model parameters or learnable components (e.g., pretraining or continued pretraining on new corpora, supervised fine-tuning, adapters/LoRA layers, vocabulary or acoustic-model updates), thereby changing the model itself. \textbf{In-context} adaptations leave parameters unchanged and instead shape behavior via inputs and orchestration (e.g., prompt and system-instruction design, few-shot exemplars, retrieval-augmented generation, tool use and pipeline routing, translation pivots, and guardrails).}

Coding proceeded in iterative cycles. We began with open coding on a subset of transcripts, then consolidated labels into a codebook that was refined through discussion and application to additional data until stable. The final codebook included 63 codes (e.g., \textit{challenges with accent}, \textit{need for good data}, \textit{challenges between modalities}), organized into 10 categories (e.g., \textit{Language Contextualization}, \textit{Social Factors}, \textit{Safety \& Validation}), providing both granularity and thematic coherence for full corpus analysis.

\parabold{Member checking} To strengthen validity, we employed member checking, returning synthesized accounts of each project to participants for review \cite{mckim_meaningful_2023}. For each project, we produced a 7–9 page summary document that outlined key findings and situated them within the emerging framework. Participants reviewed these documents, suggesting clarifications or corrections. Following qualitative best practices, we incorporated all factual adjustments, recorded points of interpretive disagreement, and revised our analysis accordingly. With only minor refinements, participants were largely in agreement with our representations.

To protect confidentiality, participants were anonymized. Raw data (audio, transcripts, notes) were stored on encrypted drives with restricted access, and any potentially identifying details were redacted from materials shared for review.

\subsection{Researcher Positionality}
We recognize that researcher backgrounds inevitably shape the design and interpretation of this work. All authors are of Indian origin. Two are researchers at a university in the United States, and two are researchers at an industry research lab in India. \added{Two authors were part of the technology development teams for \cataractbot, \ashabot, and \shiksha; however, they did not take part in the study interviews. Their experiences informed the development of the interview protocol, but all insights discussed in this paper originate from interviews conducted with 
other lead AI developers involved in the projects.} Three of the four authors have extensive experience designing and deploying technologies for education, health, and governance in non-Western contexts. This experience gave us both access to participants and deep familiarity with the institutional, cultural, and infrastructural challenges of building AI systems in these settings.

Our methodological choices were informed by this background. We used semi-structured interviews to foreground participants’ perspectives and allow them to surface challenges and tradeoffs beyond those anticipated by the research team. To reduce potential bias, all interviews were conducted by the author with limited prior professional connection to participants, helping ensure open and candid conversations. Participants were assured of anonymity, which encouraged them to share both successes and critical reflections on their projects. We see this work as part of a broader commitment to centering local expertise and documenting the opportunities and challenges that developers and domain experts experience in designing culturally aligned AI systems in high-stakes, non-Western contexts.

\section{Findings}\label{sec:findings}

\added{
Our findings show that AI developers and domain experts collaborated extensively, taking into account a range of considerations to ensure that applications worked effectively for their users. We identify six key factors, \textsc{\textbf{L}anguage}, \textsc{\textbf{I}nstitution}, \textsc{\textbf{S}afety}, \textsc{\textbf{T}ask}, \textsc{\textbf{E}nd-User Demography}, and \textsc{\textbf{D}omain} (LISTED), that shaped both system building and deployment of culturally aligned AI technologies (detailed in \S\ref{subsec:language}-\ref{subsec:domain}). To incorporate these LISTED factors, developers employed a mix of in-context and in-weight adaptations to the AI models, summarized in Table \ref{tab:project-adaptations}.
}
\newcommand{\prompt}{\incontext{\small{Prompt Engineering}}\hspace{1pt}}
\newcommand{\rag}{\incontext{\small{RAG}}\hspace{1pt}}
\newcommand{\glossaryadapt}{\incontext{\small{Glossary}}\hspace{1pt}}
\newcommand{\humanintheloop}{\incontext{\small{Human-in-the-loop}}\hspace{1pt}}
\newcommand{\modeltraining}{\inweight{\small{Model Training}}\hspace{1pt}}
\newcommand{\continualpretrain}{\inweight{\small{Continual Pre-training}}\hspace{1pt}}
\newcommand{\finetuning}{\inweight{\small{Fine Tuning}}\hspace{1pt}}
\newcommand{\instructiontuning}{\inweight{\small{Instruction Tuning}}\hspace{1pt}}

\newcommand{\projectdesc}[1]{\small{#1}}

\newcommand{\aftertags}{\\[-0.75ex]} 
\renewcommand{\tabularxcolumn}[1]{p{#1}} 
\newcolumntype{D}{>{\raggedright\arraybackslash}m{0.175\textwidth}} 
\newcolumntype{T}{>{\raggedright\arraybackslash}m{0.45\textwidth}} 
\newcolumntype{Y}{>{\raggedright\arraybackslash}X}                 

\renewcommand{\arraystretch}{1.65}

\begin{table*}[t]
\centering
\arrayrulecolor{gray!20} 
\begin{tabularx}{\textwidth}{D Y}
\arrayrulecolor{black} 
\hline
\textbf{Project} & \textbf{AI Adaptation Techniques} \\
\hline

\arrayrulecolor{gray!60}

\textbf{\vetbot} &
\rag \prompt \aftertags
\multicolumn{2}{p{0.98\textwidth}}{%
\projectdesc{A web-based veterinary assistant for farmers. It uses RAG over a curated knowledge base and carefully designed prompts to simplify responses in local dialects.}}
\\
\hline

\textbf{\farmerchat} &
\rag \prompt \glossaryadapt \finetuning \aftertags
\multicolumn{2}{p{0.98\textwidth}}{%
\projectdesc{A mobile agricultural chatbot for farmers. It combines RAG with curated knowledge bases, glossaries for local terms, and fine-tuned models to improve language and speech recognition in the agriculture domain.}}
\\
\hline

\textbf{\prompts} &
\continualpretrain \instructiontuning \humanintheloop \aftertags
\multicolumn{2}{p{0.98\textwidth}}{%
\projectdesc{An SMS-based maternal health service. It uses continually pre-trained and instruction-tuned models for low-resource African languages, with a classifier routing urgent queries to human experts.}}
\\
\hline

\textbf{\cataractbot} &
\rag \humanintheloop \prompt \aftertags
\multicolumn{2}{p{0.98\textwidth}}{%
\projectdesc{A WhatsApp-based bot supporting patients before and after eye surgery. It uses RAG with vetted clinical and logistical resources, with human doctors reviewing and revising responses before delivery.}}
\\
\hline

\textbf{\ashabot} &
\rag \humanintheloop \prompt \glossaryadapt \aftertags
\multicolumn{2}{p{0.98\textwidth}}{%
\projectdesc{A WhatsApp-based bot for community health workers. It combines RAG, glossary adaptations for local terms, and prompts for a friendly tone, with unanswered queries escalated to human experts.}}
\\
\hline

\textbf{\padhai} &
\modeltraining \humanintheloop \aftertags
\multicolumn{2}{p{0.98\textwidth}}{%
\projectdesc{A mobile child reading assessment tool. It trains ASR models on children’s speech and integrates human review to correct outputs and improve accuracy over time.}}
\\
\hline

\textbf{\shiksha} &
\rag \instructiontuning \glossaryadapt \humanintheloop \prompt \aftertags
\multicolumn{2}{p{0.98\textwidth}}{%
\projectdesc{A web-based lesson-plan generator for teachers. It uses RAG to ground lesson content in official curricula, instruction-tuned models for vernacular translations, and teacher reviews for classroom adaptation.}}
\\
\hline

\textbf{\suvas} &
\finetuning \glossaryadapt \humanintheloop \aftertags
\multicolumn{2}{p{0.98\textwidth}}{%
\projectdesc{A web-based legal translation system. It relies on fine-tuned NMTs for legal vocabulary, with human translators post-editing and judges verifying outputs.}}
\\
\hline

\end{tabularx}
\par\medskip
\noindent
\inweight{\small{In-weight}} \incontext{\small{In-context}}
\par\medskip
\caption{AI adaptation techniques used across projects. The explanations of these techniques are provided in Appendix \ref{appendix:adaptation-techniques}.}
\label{tab:project-adaptations}
\end{table*}

\subsection{Language} \label{subsec:language}
AI developers and domain experts collaborated to ensure that the AI solutions function effectively in low-resource and non-Western languages. While language also encompasses facets such as domain-, geography-, and culture-specific vocabulary, these are discussed in detail in subsequent sections. Here, we focus on the \emph{basic language capabilities} of the systems --- whether they can correctly understand input and produce coherent output in the target low-resource languages. Even such capabilities required dedicated efforts, as these languages are often poorly supported in mainstream LLM models~\cite{ojo_afrobench_2025, dargis_evaluating_2024, singh_global_2025}.

\subsubsection{Language and Technical Considerations} \label{subsub:model-selection}
Developers made key decisions about which models best handled target languages, balancing fluency, domain accuracy, and model availability.
\parabold{Selecting most suitable models} The AI developers began by identifying models best suited to their target language and context. They worked with domain experts to ensure that the linguistic nuances of the target users are accurately captured in the model. Because commercial and open-source models differed widely in language capabilities, the teams tested several configurations before deciding which to deploy. For example, in \shiksha, GPT-4 generated lesson plan content in English, while the Indic model Sarvam 2B translated it into Kannada. Although the lesson plans were meant for Kannada, they were first created in English because GPT-4 produced pedagogically high-quality content in English but struggled in Kannada. Sarvam 2B, by contrast, generated fluent Kannada but with weaker pedagogical quality. To reconcile these trade-offs, developers adopted a hybrid pipeline, assigning each model to the task it handled best.

Most projects relied on informal, qualitative assessments to evaluate language capabilities, but \farmerchat adopted a more systematic approach. Working with domain experts, they fine-tuned a Llama 3 model with agricultural vocabulary to identify domain-specific terms across languages. They also built gold-standard transcripts for multilingual voice datasets collected from their users. Each time a new ASR model was released, they evaluated it by comparing ASR outputs with these transcripts using the fine-tuned Llama model. Performance was measured with standard metrics such as word error rate and an “agree score” that penalized mistakes on agricultural terms — crop names, pests, diseases, units, numerals — which accounted for about 20\% of the vocabulary used by the farmers in \farmerchat. Even a single error (e.g., mistaking \textit{mushroom} for \textit{masoor(red lentil)}) could distort a farmer’s query. This evaluation process revealed substantial variation across ASR models and allowed \farmerchat to select the best available model for each target language.
\added{
\parabold{Geographical disparities in model support} Developers working across multiple regions identified stark geographical disparities in available models. This disparity was especially pronounced for teams like \farmerchat working across nationalities. As \farmerchatdomain explained:

\begin{quote}
\textit{``In India, even though it’s not perfect, at least many of the major languages are supported by the models. But in Africa it’s much harder, because a lot of the languages just aren’t supported even though there are so many users. For example, in Ethiopia we work in two zones that use two different languages, Amharic and Oromo. Amharic is supported to some extent, but Oromo isn’t supported at all in most models.''} (\farmerchatdomain)
\end{quote}
}

\parabold{Temporality of model selection} The AI developers emphasized that model selection was not a one-time technical decision but an ongoing process shaped by both language requirements and the evolving availability of technologies at different points in time. For example, \suvas{}’s use of NMT to translate legal documents from English to vernacular languages was shaped by the technologies available to developers at the time. As one \suvas developer reflected:

\begin{quote}
    \textit{``\suvas was developed in 2019 when using LLM-based approaches was not yet possible. This led us to adopt NMT as the default.''} (\suvasaidev)
\end{quote}

These experiences underscore that effective multilingual AI deployment requires treating model selection as a dynamic, iterative process rather than a static technical specification.

\subsubsection{Human Infrastructure}\label{language-variation-modality}
Effective language support depended not only on models but on human labor, as domain experts and field staff identified shortcomings, created workarounds, and supplied the resources needed to train or adapt systems.
\parabold{Role of domain experts} Given the uneven language support in the AI landscape — where certain languages and dialects are better supported than others — the human effort required to contextualize systems varied across both language and modality (text or voice). Domain experts contributed in two main ways: (1) identifying the language shortcomings of the systems and (2) improving their language capabilities.

Language shortcomings were often surfaced through feedback mechanisms from domain experts. All projects conducted regular log analyses and used multiple channels to gather user feedback. For instance, in \ashabot and \cataractbot, domain experts received daily logs of interactions by email, enabling them to spot system errors and ask AI developers to apply fixes where possible. In addition, field staff collected feedback through their on-ground programmatic activities. In \vetbot, for example, field volunteers not only trained farmers to use the application but also relayed user feedback directly to the developers. 

\parabold{Resource-based adaptation strategies} Acting on this feedback, however, depended on organizational resources. Most teams began with prompt engineering because it was inexpensive. When \vetbot{}’s staff reported that Tamil Nadu farmers used dialectal variants the system failed to recognize — for example, different words for \textit{``fever''} — developers tried adding these variants to prompts. But as cases multiplied, the approach proved unsustainable, and users were asked to communicate in ``standard Tamil.'' Better-resourced teams pursued more robust fixes. \farmerchat, for example, supported Bhojpuri-speaking farmers in Bihar by co-developing a glossary of Bhojpuri words commonly substituted for Hindi. Because models handled Hindi far better than Bhojpuri, the glossary allowed the system to map terms into Hindi before processing — an easier solution than training a Bhojpuri model from scratch.

\parabold{Human effort for in-weight approaches} When in-context techniques proved ineffective, developers relied on in-weight approaches such as pre-training and fine-tuning, which were far more resource-intensive. \padhai required STT models for conducting reading assessments in Hindi and Marathi. Existing STT models, designed for adult speech, performed poorly with children, prompting \padhai{}’s parent organization to build a new model. With support from field staff and partner organizations, they collected $\sim$2,500 hours of voice data from 180,000 children. Annotation was similarly labor-intensive: in addition to trained annotators, 2,900 volunteers contributed by annotating 33,000 samples, spending a total of 475 hours.

Unlike Tamil or Bhojpuri, where prompt engineering or glossaries could mitigate challenges, certain very low-resource African languages such as Hausa and Xhosa lacked even minimal model support. To fill this gap, \prompts undertook continual pre-training and instruction tuning of their own LLM --- a resource-intensive effort that demanded close collaboration with field partners to annotate training data. Despite the cost and difficulty, the team acted quickly, recognizing the urgency of reaching users otherwise excluded. As \promptsaidev reflected:
\begin{quote}
\textit{``If I waited a year for the GPT models to be good enough, that's hundreds of thousands of moms that may not have had their questions answered.''} (\promptsaidev)
\end{quote}

These examples demonstrate that the human resources needed to accommodate various languages and communication formats differed significantly. The approaches ranged from low-effort methods like prompt engineering to highly resource-intensive strategies such as large-scale data collection and model training, depending on both the organization's available resources and the specific linguistic complexities involved.

\subsection{Institution}\label{subsec:institution}
Institutional factors enabled broader systemic adoption. These involved aligning systems with government regulations and policies, organizational workflows, and institutional priorities. 

\subsubsection{Aligning with Institutional Requirements and Knowledge}\label{subsub:institutional-requirement}
Institutional alignment was essential for adoption, requiring systems to comply with policies, embed institutional knowledge, and safeguard user privacy.

\parabold{Policy Compliance} For institutional adoption, alignment with institutional requirements was essential: compliance with rules and embedding institutional knowledge such as curricula in education, medical protocols in health, or legal terminology in courts. Domain experts played a critical role in surfacing these requirements, and in working with AI developers to embed them into the system’s outputs. For example, in \shiksha, lesson plans had to be created in formats specified by government guidelines for teachers. As \shikshadomain explained, the format itself was crucial: regardless of the quality of the content, teachers would not consider using AI-generated lesson plans unless they matched the required structure, since the documents also served as official records for submission to their supervisors. Domain experts were critical in interpreting such requirements, which varied across different states. For example, Telangana mandated a different lesson plan format than Karnataka. Comparable requirements arose in \farmerchat, where institutional requirements shaped how the system had to be adapted. Operating across multiple states and countries meant that its advice had to align with the guidelines of local agriculture departments, which varied by region and were binding on farmers. As \farmerchatdomain explained:

\begin{quote}
\textit{``If you work in Andhra Pradesh, the government there doesn't want to promote chemical practices at all. They only want regenerative agriculture, natural farming, etc. For that particular location, in the database, it's only their documents. We would never even call the general knowledge base.''} (\farmerchatdomain)
\end{quote}

\parabold{Institutional Knowledge} To ensure compliance, AI developers relied on domain experts to curate knowledge base resources that accurately reflected governmental policies, as these experts were closely connected with the relevant government bodies.  For instance, in \ashabot, patients often asked health workers about government schemes, and health workers depended on \ashabot for accurate answers. When Rajasthan introduced the LADO scheme, which provides financial incentives to families with a baby girl, the system initially failed to respond because the policy was absent from the knowledge base. Domain experts identified this gap through user feedback and log analysis, then updated the knowledge base so future queries could be answered correctly.

\parabold{Privacy Compliance} Privacy was another critical concern given the data-intensive nature of these applications.
For in-weight adaptations, organizations had to comply with institutional data protection requirements. For example, \padhai follows India's Digital Personal Data Protection Act, 2023, which mandates lawful and transparent processing of personal data and requires consent from parents (or guardians) in the case of minors. Accordingly, \padhai obtains parental consent for the use of children’s voice data. In \cataractbot, all personally identifiable information (PII) is removed before queries are sent to the LLM, while non-PII attributes such as age and gender are retained to generate more contextualized responses.

Collectively, these instances show that institutional alignment was as critical as technical performance. From formats and policies to privacy safeguards, systems could gain legitimacy and adoption only when they embedded institutional rules and practices directly into their design.

\subsubsection{Institutional Support}\label{subsub:institutional-support}

Support from institutions, including government agencies, deployment partners, and field staff, played a central role in determining how AI systems moved from development to real-world use. Institutional backing shaped the visibility of these systems, influenced user trust, and ultimately affected patterns of adoption. These forms of support were themselves tied to the institutional capacity available within each country.
\added{
}
\added{
\parabold{Institutional Capacity}
Institutional capacity for developing AI systems depends on both organizational resources and the broader environments in which organizations operate. Some regions benefit from strong government support and established research infrastructures, while others lack them. This produces a geography of uneven technological production in which well-resourced contexts develop new systems and others rely on adapting tools created elsewhere. The trajectories of \padhai and \suvas illustrate this pattern. Both were created by Indian organizations for local users, and their use in Colombia and Bangladesh has centered on adaptation rather than building new systems. These inter-dependencies are reinforced by the global concentration of AI expertise and infrastructure, including widespread reliance on foundation models developed by large US-based companies. As \promptsaidev, an African developer, noted, many NGOs in resource-constrained settings cannot match the salaries or working conditions of major technology firms, limiting their ability to recruit AI specialists. AI therefore amplifies existing institutional disparities because the capacity to develop and sustain such systems remains unevenly distributed across regions.
}

\parabold{On-ground Support} In many cases, such as \shiksha, the institutional support was mobilized not only through formal structures but also through the broader excitement surrounding AI. \shikshadomaintwo explained how the launch of the AI-driven lesson planning tool by the Education Minister significantly boosted its visibility, encouraging more teachers to adopt it. In addition, integrating \shiksha in the existing structures of the partner organization, such as training and ongoing support from teacher mentors, resulted in more adoption. Similar dynamics played out in health context. In \cataractbot, hospital counselors were central to onboarding patients, including guiding them on how to use the application. On-the-ground support also proved indispensable in \ashabot, where early assumptions about WhatsApp’s familiarity proved misplaced. As \ashabotdomain reflected:

\begin{quote}
\textit{``Initially, we assumed that since health workers already used WhatsApp, no training on the bot was necessary. Over time, we realized training was essential, given the variation in their digital literacy, so in the new district rollout, we now provide it.''} (P12)
\end{quote}

\parabold{Institutional Trust} Trust-building was another dimension where institutional partnerships played a decisive role. In \prompts, deployments began only after formal MOUs with county governments were signed. The service was branded under the county’s name, rather than as \prompts, allowing mothers to engage with it as a trusted government service. In contrast, \farmerchat faced greater challenges in building trust. When the system was advertised on Facebook, many potential users were skeptical, assuming that free services must be scams. As \farmerchatdomain noted, overcoming this skepticism requires time and the gradual building of a critical mass of users, whose participation in turn inspires trust for others to join. Yet, despite these challenges, Facebook ads remained necessary for scale, as in-person onboarding of farmers was too resource- and time-intensive.

\parabold{Challenges without institutional support} The absence of institutional support was most stark in \vetbot. Attempting to scale the AI tool through volunteers, without government or formal institutional involvement, proved to be a major bottleneck. Unlike other cases, the lack of structured institutional involvement hindered adoption and limited its reach. 

These findings show that institutional support was not just helpful but critical. AI systems could scale only at the speed of institutional trust. Without such backing, even well-designed systems struggled to gain traction.

\subsection{Safety}\label{subsec:safety}
Safety factors were central across the design, implementation, and evaluation of these systems, with an emphasis on ensuring outputs were reliable and did not cause harm to end-users. \added{While core concerns such as accuracy and prevention of overtly risky behaviors were systematically addressed, more subtle forms of harm — such as cultural, caste, and identity-based biases — received no attention due to limited time and resources.} Safeguards were implemented at multiple points in the workflow, most notably at two stages: before inputs entered the system and during the processing of outputs.

\subsubsection{Safety Measures Before Processing Inputs}\label{subsub:mesaures-before-processing}
Safety safeguards began even before inputs reached the main AI pipelines, combining curated knowledge, prompt guardrails, and embedded classifiers to prevent harmful or unlawful outputs.

\parabold{In-context Techniques} In systems that relied on knowledge bases, as described in \S\ref{reliance-on-curated-knowledge}, domain experts played a key role in guaranteeing safety by curating knowledge artifacts only from verified and authoritative sources. This process ensured that the foundations of the AI systems were trustworthy before any responses were generated. AI developers also introduced technical safeguards at the input stage. A basic measure was to use prompt guardrails to filter harmful or irrelevant queries. In \ashabot, queries related to a baby’s gender reveal are rejected, since revealing a baby’s gender during pregnancy is prohibited in India.

\parabold{In-weight Techniques} In \prompts, safety mechanisms were embedded directly into the model through continual pre-training and instruction-tuning. The team incorporated the BeaverTails safety dataset \cite{ji_beavertails_2023}, which contains adversarial prompts designed to expose unsafe behaviors, making the system more robust. Beyond this, they developed a custom input classifier to filter messages before they reached the LLM. This required substantial human effort: roughly 100,000 messages sent by users to the human helpdesk were manually annotated for training data. A RoBERTa-based model was then trained to classify message intent and assign an emergency-risk score, which determined whether a query was routed to the LLM or escalated to the human helpdesk (\S\ref{human-in-the-loop}). As \promptsaidev noted, this classifier was particularly important because incoming messages were highly code-mixed, often blending vernacular languages with English — an area where off-the-shelf intent classifiers consistently failed. The broader challenges of ensuring safety in multilingual settings were underscored by \aiexpertone:

\begin{quote}
\textit{``Security vulnerabilities are hard to manage in multilingual contexts. Models are trained with guardrails to block dangerous requests, like instructions for weapons or content about self-harm, but these protections can break down in other languages. That creates two risks: malicious users may exploit the gap, and ordinary users may unintentionally trigger unsafe outputs when discussing sensitive topics the model does not recognize. The problem is especially visible in areas like suicide prevention, where cultural differences and metaphorical language make detection more difficult.''}(\aiexpertone)
\end{quote}

These strategies highlight that ensuring safety required not only technical guardrails but also substantial human effort, especially in multilingual settings where risks were harder to detect.

\subsubsection{Measures During Workflow Processing}\label{subsub:mesaures-during-processing}

Both machine- and human-driven measures were employed during data processing to ensure safety, with human oversight often taking precedence over automated checks.

\parabold{Machine Oversight} A common safeguard was the use of LLM validation agents to audit outputs before delivery. This strategy was implemented in applications such as \shiksha, \farmerchat, and \prompts. The validation agents were guided with examples and instructions to review outputs and reject those that did not comply with established guidelines. In \prompts, the validation agent checked the accuracy of responses against the mother’s question before releasing them.

\parabold{Human Oversight} Human oversight remained the cornerstone of safety, with technical safeguards serving as complements rather than substitutes. The balance between automation and human review depended on the stakes of the task and the maturity of the models. In \prompts, non-medical queries were handled by LLMs, while medical queries were routed directly to human experts. Oversight was especially strict in new languages: while mature Swahili models could deliver responses directly, Hausa outputs were reviewed by a helpdesk expert until developers gained confidence through feedback. In \cataractbot, every message was reviewed by doctors before being sent to patients, with verified responses marked by a badge to reinforce trust. Other domains embedded human oversight into task workflows. In \padhai, field evaluators could override or correct model outputs. Patterns of errors flagged by evaluators were then fed back to improve the model. In legal contexts, \suvas combined AI assistance with rigorous human validation. Expert translators produced drafts with the system, but final documents required judicial approval before release. Even then, judgments carried disclaimers noting that the translations were for general information only and had no legal standing.

These workflows show that reliability depended less on automation alone than on carefully balanced human–machine oversight tailored to the risks of each domain.

\subsection{Task}\label{subsec:task}

Our participants also had to account for task-specific factors, including the physical environments in which applications were used, model design choices to meet task demands, and latency requirements.

\subsubsection{Environmental Requirements}\label{subsub:task-environement}
AI developers emphasized the importance of investing effort into understanding the environment within which an AI application would be used. In \padhai, for example, AI developers and domain experts initially collected children’s voice data using noise-canceling microphones to ensure high-quality recordings. However, once the model was deployed in real-world settings, they realized that the typically environments where the assessments are done are not noise-free. The model trained on noise-free data performed poorly under those conditions, requiring the team to redo their data collection to reflect these real-world conditions. A similar adjustment shaped \farmerchat{}’s design. Because farmers primarily interact with the application while working in the fields, voice data was collected in those same farming environments to ensure that the models could handle the background noise typical of such settings.

\subsubsection{Task Requirements}\label{subsub:task-specific-needs}
Different tasks demanded distinct technical choices, from preserving student errors to producing nuanced translations or adapting tone and style to application needs.

\parabold{Task dictates design} The goal of a task determined the design of underlying AI systems. For instance, for most ASR systems, the aim is to interpret the intended meaning of speech. But in \padhai, the purpose was different: to capture students’ mistakes with precision. This required a fundamental shift in design — rather than relying on language models to smooth errors, as standard ASR does, the system emphasized the acoustic signal so that mistakes were preserved instead of corrected.

\begin{quote}
\textit{Typical ASR systems correct user errors by relying heavily on language models. In assessment tasks this is harmful because mistakes must be preserved, so the system should depend more on the acoustic signal than on language modeling.} (\padhaidomain)
\end{quote}

\parabold{Translation requirements} Task requirements also shaped translation strategies. In \shiksha, the challenge was generating Kannada lesson plans from English. Commercial NMT systems like Azure AI Translate handled short sentences reasonably well but struggled with long lesson plans, where translations needed to be less mechanistic and more contextually nuanced. To address this, the team instruction-tuned an Indic LLM with long academic paragraphs, enabling it to manage long-form academic content translation.

\parabold{Task-driven stylistic choices} Other applications faced equally task-specific but stylistically distinct demands. In \ashabot, the goal was to sustain an informal, conversational tone that mirrored how health workers spoke with patients — an intent reflected in the WhatsApp bot’s name, \textit{Saheli} (``female friend'' in Hindi), which signaled approachability. By contrast, \suvas required precise and standardized legal terminology for court translations. Although legal documents are long and complex — making them well suited to LLMs — the team prioritized vocabulary control over flexibility. NMTs offered this control, so despite the advantages of LLMs for handling long-form text, the system continued to rely on NMTs. These examples illustrate how translation and language-processing choices varied across applications depending on task requirements, whether accuracy of speech errors, handling of long-form content, conversational tone, or strict adherence to legal vocabulary.

\subsubsection{Latency Needs} \label{cost-latency}

Latency — the delay between input and response — was central to usability, pushing developers to weigh quality, cost, and responsiveness.

\parabold{Critical in realtime chats} Latency was especially important in applications like \farmerchat, where farmers expected real-time conversational exchanges. As \farmerchataidev explained, fine-tuned LLMs generated more domain-aligned responses but were costly to train and serve. To balance quality and cost, the team experimented with LoRA adapters, which preserved many of the quality gains at lower cost. However, this approach introduced significant inference latency, slowing down responses to the point that it undermined the user experience and proved unsuitable for deployment.

\parabold{More control with in-weight approaches} Applications that carried out extensive in-weight adaptations had greater control over latency. In \padhai, reading assessors conducted large numbers of assessments daily and could not afford delays, especially while working with children. To address this, the system was designed to run models offline directly on assessors’ devices, substantially reducing latency and ensuring functionality even in areas with poor or no connectivity. Similarly, \prompts avoided latency challenges by relying on instruction-tuned models that generated responses directly, without extra steps such as translation or retrieval.

\parabold{UX-based mitigation} Systems that relied more heavily on in-context adaptations experienced considerable latency delays. In \ashabot and \cataractbot, multiple steps — translation before and after RAG workflows, and in some cases human verification — added to latency. Developers adopted expectation management and UX workarounds to mitigate frustration. In \ashabot, health workers were asked to pause at least 30 seconds before sending the next message. In \cataractbot, patients were informed upfront that responses would arrive only after a doctor’s review. These strategies helped normalize delays but could not eliminate them, leaving responsiveness constrained in time-sensitive interactions.

These cases show that managing latency required both technical adaptations and user-facing adjustments, making latency a design choice as much as a performance constraint.

\added{\subsection{End-User Demography}\label{subsec:demography}}
End-User Demographic factors — such as geography, culture, age, gender, and income — were also critical in shaping how end-users engaged with the applications. In particular, aligning systems with cultural symbols and geographic norms was a pervasive concern, as regional variation directly influenced their relevance and adoption.

\subsubsection{Geography and Culture}\label{subsub:granularity-challenge}
\added{Geographic and cultural diversity demanded careful adaptation, as systems struggled to balance broad regional coverage with hyper-local variation that shaped everyday relevance.}
\parabold{Western-centric defaults} Our participants noted that geographical diversity of the users introduced multiple layers or granularity: across geographic states (different languages), within states (regional variations), and even within regions (dialectical differences). Each layer shaped not just the vocabulary used but also the cultural references and knowledge embedded in applications. Addressing this granularity required deliberate effort, as AI models often defaulted to Western-centric norms and practices. This was especially salient in \shiksha, where lesson content had to match learners’ environments. \shikshaaidev explained, LLMs often produced Western examples less relevant for Indian learners:

\begin{quote}
\textit{``For example, when teaching a chapter on components of food and trying to show the vitamins or minerals present in dishes that students commonly eat at home, the AI often produces examples like Caesar salad because it is trained on Western data. Indian children, especially in rural areas, are not exposed to such food. This makes it harder for them to understand the context of the lesson, which in turn might reduce the learning outcomes.''} (\shikshaaidev)
\end{quote}

To mitigate this, \shiksha relied on prompt engineering to explicitly request culturally relevant Indian examples. In contrast, other projects with domain- and application-specific knowledge bases relied on selective retrieval through RAG to handle regional variation and cultural adaptation. For instance, in \cataractbot, the food that the patients consumed after surgery varied based on regions. The system had to retrieve the correct regional knowledge chunks to generate accurate responses. While higher-level adaptations were generally easier to achieve using these approaches, fine-grained contextualization remained nearly impossible at scale, since communities often use hyper-local terms and explanations intelligible only within small geographic pockets. \farmerchatdomain explained:

\begin{quote}
\textit{``I’m sure this is not just in agriculture, it is also in health and other domains. Local communities everywhere have very specific ways of explaining things and use words that are understood only within that small geographic area. These variations change a lot, and this kind of contextualization is almost impossible for an large-scale application like \farmerchat to manage, at least for now.''} (\farmerchatdomain)
\end{quote}

\parabold{Regional Clustering} This challenge led to a pragmatic compromise: clustering fine-grained localities into broader regional categories that preserved meaningful differences while remaining computationally feasible. In Karnataka, \shiksha{}'s domain experts identified three regional clusters — North, South, and Coastal — that captured the most educationally relevant variations in dialect, cuisine, climate, and local flora and fauna. In every lesson plan, rather than attempting comprehensive local coverage, the AI system generated multiple regional examples, delegating final selection to teachers' local knowledge.

Data collection efforts for in-weight adaptations also used a similar strategy, prioritizing representativeness over exhaustive coverage. As \padhaidomain explained:

\begin{quote}
\textit{``It's not possible to cover the entire state. But of course, within these states, we chose districts such that they covered a wide variety of linguistic diversity. We used the Linguistic Diversity Index, which signifies how many different mother tongues are spoken in a particular region. So, based on that we tried to optimize for the most diversity. For example, in Uttar Pradesh we selected 10 districts that together reflected the maximum possible diversity. 
''}
(\padhaidomain)
\end{quote}

\parabold{Geography-specific Glossaries} \added{Developers also experimented with geography-specific glossaries to address local variation. Just as a Bhojpuri glossary supported farmers in Bihar, \farmerchat has been curating glossaries for different regions, while \suvasdomain stressed the importance of partnering with local experts and government agencies to build such resources.} 

These approaches reveal a fundamental tension in AI localization: the impossibility of capturing fine-grained local variation within scalable systems. The projects managed this through strategic abstraction — identifying the level of geographic granularity that balanced cultural relevance with operational feasibility. Rather than pursuing exhaustive local coverage, systems created manageable regional clusters, built representative datasets using diversity indices, and relied on local experts to bridge the final gap between regional categories and hyper-local contexts. This division of labor emerged as a pragmatic necessity, with AI handling broader regional variations while human expertise provided the fine-grained cultural translation that no system could anticipate or encode at scale.

\subsubsection{Other Demographic Factors}\label{subsub:other-demography}

Factors such as age, literacy, income, and gender shaped AI system design across three key areas: interaction modalities, access mediums, and underlying content.

\parabold{Interaction Modalities}
Preferences for text or voice were shaped largely by literacy, age, and accessibility. In \farmerchat, \vetbot, and \ashabot, voice was essential because users were far more comfortable speaking than typing. This created technical hurdles, since mainstream ASR systems offered limited or no support for many low-resource languages. Text support in AI models did not imply voice support: for instance, LLMs could generate Kikuyu text for \farmerchat's users, but no Kikuyu ASR existed. Age and accessibility further reinforced demand for voice. In \cataractbot, elderly patients preferred voice notes due to visual impairments and familiarity with spoken communication. Even where voice tools existed, quality and cost shaped implementation. \vetbot initially used Chrome’s free STT, but users said its Tamil output sounded like \textit{“foreign Tamil spoken on BBC Radio”}, leading developers to switch to Google’s paid Speech AI for more natural interaction.


\parabold{Access Mediums} The medium through which the user accessed the tool was driven by factors such as income and infrastructure considerations. \prompts was deliberately delivered as an SMS service so that low-income mothers with only feature phones could access it without mobile data costs. In contrast, many Indian chatbots leveraged WhatsApp, where data was relatively inexpensive and adoption widespread. However, platform choices involved complex tradeoffs beyond user preferences. While WhatsApp offered user-friendly interfaces, using the WhatsApp Business API required lengthy logistical processes to set up accounts with Meta. These constraints forced \vetbot to deploy via a web app despite WhatsApp being more accessible to users. Similarly, \shiksha shifted from a chatbot-based service to a web app primarily because teachers required richer interfaces to edit lesson plans — functionality that simple messaging platforms couldn't support.

\parabold{Content Considerations} Demographic factors also shaped the underlying data and content within systems themselves. Gender influenced linguistic patterns: in \farmerchat, women farmers often used non-technical terminology to describe agricultural machinery, while men relied more on technical terms. Age created similar vocabulary differences, with older farmers engaging systems using traditional agricultural terminology and practices, while younger farmers were more open to exploring modern techniques. These demographic differences required deliberate attention to data collection and model training. In \padhai, developers made conscious efforts to ensure gender diversity in children's voice data collection, recognizing that voice models are sensitive to gender. Resource constraints also influenced content design itself. \shiksha deliberately designed lesson plans containing activities that required minimal or no materials, making them usable in under-resourced classrooms. Similarly, \vetbot ensured its recommendations were realistic and actionable for low-income farmers rather than suggesting expensive solutions beyond users' means.

These examples demonstrate that demographic factors shaped every layer of system design — from modalities and platforms to content and interaction styles. Addressing literacy, age, income, and gender considerations was not an afterthought but fundamental to creating AI systems that could function effectively in diverse, resource-constrained contexts.

\subsection{Domain}\label{subsec:domain}
Domain factors ensured that AI systems could meet the complex demands of specialized tasks by incorporating the necessary domain knowledge and vocabulary needed. While curated glossaries helped with terminology, capturing and applying deeper expertise still required significant involvement from domain experts, as discussed below.

\subsubsection{Curated Knowledge Bases over LLM World Knowledge}\label{reliance-on-curated-knowledge}

In high-stakes domains, AI systems relied on curated knowledge bases to ground outputs in verified data, reducing risks from LLM world knowledge while ensuring domain and context specificity.

\parabold{Knowledge grounding for domain specificity} Developers emphasized the central role of knowledge bases in producing context-sensitive, domain-specific outputs. In high-stakes areas such as healthcare, responses had to be grounded in verified sources rather than an LLM’s world knowledge. Most projects relied on Retrieval-Augmented Generation (RAG) \cite{NEURIPS2020_6b493230}, using expert-curated resources as the backbone of the applications. The extent to which developers used an LLM’s world knowledge varied with the stakes. In \shiksha, lesson plans were anchored in prescribed textbooks, with LLM knowledge used only for supplementary activities. Even so, substantial human oversight was required: teachers had to review roughly 800 lesson plans per grade across six grades in every state of deployment. This review must be repeated whenever curricula change, creating recurring and unpredictable workloads. As \shikshadomain noted, this ongoing effort threatens the project’s long-term sustainability, especially as it expands to new regions.

In health applications such as \ashabot, \cataractbot, and \vetbot, generation was fully grounded in verified knowledge bases maintained by domain experts. Their involvement was essential not only for curation but also for catching retrieval errors. In \vetbot, for example, experts found that the system returned treatments for goats when asked about lambs, due to subtle distinctions in Tamil (\textit{aadu} for goat and \textit{semmari aadu} for lamb). Such cases highlight the challenges RAG faces in low-resource languages, where small lexical nuances can lead to serious errors.

\parabold{RAG systems in English} To mitigate such risks, several projects chose to build their RAG knowledge bases in English, even though the end users interacted with the systems in vernacular languages. As a result, any local-language knowledge artifact had to be translated before it could be added to the RAG system. For example, \farmerchat's parent organization has disseminated agricultural knowledge through thousands of farmer-created videos since 2008. The transcripts of these videos also inform the knowledge base in the RAG system. Because these transcripts were in local languages, they could not be used directly. Although AI developers implemented machine translation to reduce effort, substantial human oversight was still required to ensure accuracy.

\parabold{Building upon pre-existing data} A key enabler for applications such as \farmerchat and \prompts was the availability of rich pre-existing data, which allowed the use of in-weight adaptation techniques such as fine-tuning and instruction tuning to develop domain-specific models. As \promptsaidev emphasized, the main challenge was not the technology but the data, noting that \textit{``having good, local, up-to-date contextual data is essential''} and explaining how they receive thousands of questions daily from mothers to expand and update their databases, ensuring models can be refreshed with the most recent data every six months or annually. Similarly, \farmerchataidev noted that maintaining fine-tuned models across languages was less a technological challenge than a data one --- the real difficulty lay in the human effort required for preparing and managing high-quality datasets:
\begin{quote}
\textit{``Fine-tuning itself is not the difficult part. Around 80\% of the effort goes into preparing and collecting the data. Once the architecture, framework, and code are in place, I can adjust hyperparameters and fine-tune another model with relative ease.''} (\farmerchataidev)
\end{quote}

These cases highlight that while knowledge bases were the backbone of effective AI applications, their success depended less on the sophistication of the underlying models than on the availability, quality, and continual curation of domain- and context-specific data.

\subsubsection{Human-in-the-Loop Workflows}\label{human-in-the-loop}
Human-in-the-loop (HIL) workflows enabled expert intervention whenever AI systems fell short, ensuring appropriate oversight of domain risks, responsiveness to timely needs, and alignment with local contexts.

\parabold{Human Fallbacks} \added{Even with curated knowledge bases and fine-tuned models, domain experts remained indispensable during use. Participants noted that chatbots inevitably encountered questions outside their knowledge bases, making HIL workflows essential for filling gaps and ensuring accuracy.} In \cataractbot, when the knowledge base lacked an answer, queries were routed to the primary doctor. Although responses were sometimes delayed, patients expected their doctors to be busy, and the information still reached them directly.
By contrast, the same approach struggled in \ashabot. When health workers posed questions not covered in the knowledge base, the system forwarded them to local knowledge experts via WhatsApp. Multiple responses were then crowdsourced and synthesized by the LLM into a single answer. Crowdsourcing was important because the local knowledge experts were not trained doctors and might have gaps in their knowledge. While this reduced individual errors, it often failed in practice: health workers usually needed immediate answers to respond to patients, but local experts — occupied with their own responsibilities — frequently replied too late. By then, health workers had either sought information elsewhere or, in some cases, left the patient without guidance. The contrast between \cataractbot and \ashabot reveals how timing requirements and expert capacity determine the viability of HIL approaches.

\parabold{Delegation based on risk} \prompts adopted a different HIL model: the AI system handled non-urgent and non-medical queries, while medical and urgent queries were routed to a dedicated help desk staffed by human agents. This hybrid division of labor was shaped by two factors: (1) their use of an instruction-tuned LLM, which generated responses with greater variability and less predictability compared to the more controlled outputs of a RAG-based approach, and (2) the availability of trained staff who could provide mothers with more personalized, context-sensitive support. This division of labor succeeded because it matched system capabilities to query types while ensuring dedicated human capacity for high-stakes interactions.

\parabold{Accountability considerations} In domains where errors carried legal consequences, HIL workflows shifted toward human-led rather than AI-led processes. In \suvas, which was used for translating legal documents in courts, expert court translators worked in a HIL role. They used the AI system to generate initial drafts but performed most of the work themselves, ensuring that legal vocabulary was precise and consistent. Given the official status of these documents, ultimate responsibility for accuracy and validity rested with the human translators rather than the AI.

\parabold{Layered HIL processes} Educational contexts presented a different challenge: the need for multiple layers of human judgment to address both pedagogical quality and local adaptation. In \shiksha, lesson plan preparation followed a layered HIL process. First, a group of teachers used the AI system to curate plans — correcting language errors, ensuring pedagogical quality, and aligning content with state-level curricula. These curated plans were then passed to classroom teachers, who conducted a final round of contextualization with AI support. They adapted the plans to their students and local geography, making hyper-local adjustments that could not be anticipated earlier. This step was especially important given the difficulty of capturing fine-grained geographical variations in advance, as discussed in \S\ref{subsub:granularity-challenge}.

The varied approaches across these deployments demonstrate that effective HIL workflows required careful alignment between domain needs, expert capacity, and system limitations. Success depended not on simply adding human oversight, but on designing intervention points that matched domain-specific requirements — whether ensuring medical safety, legal accountability, or educational relevance. Human expertise thus served as the critical bridge between AI capabilities and real-world deployment constraints.

\section{Discussion}\label{sec:discussion}

\added{
Our findings identify six cross-cutting factors that shaped how culturally aligned AI systems are built and deployed in high-stakes, non-Western contexts (Figure \ref{fig:factors_and_influences}). \textbf{Language} required grounding in local languages through technical adaptations and community-driven resources. \textbf{Institutions} acted as mobilizers, with formal rules and informal legitimacy determining scale. \textbf{Safety} provided the trust infrastructure through indispensable human oversight complemented by technical safeguards. \textbf{Tasks} imposed constraints (latency, text length, environmental conditions) forcing departures from standard practices. \textbf{End-User Demography} shaped relevance through literacy, age, gender norms, and access channels. \textbf{Domain} considerations anchored reliability by establishing what counted as correct or trustworthy, necessitating sustained domain expert involvement.

\textbf{The LISTED factors did not operate in isolation but interacted to make AI usable, legitimate, and trustworthy.} Language, Domain, and End-User Demography grounded systems in local realities, while Institution, Task, and Safety determined scalability and reliability. Literacy constraints (End-User Demography) intersected with latency requirements (Task) to drive voice-first design; institutional mandates on formats and protocols (Institution) reinforced domain experts' role in safeguarding accuracy (Safety). Building AI systems for social good thus requires balancing multiple, interdependent constraints rather than optimizing individual factors.

While some of these factors apply to any software project, AI’s stochastic and data-dependent behavior \cite{bender_dangers_2021} introduces distinct complexities, as we show in the following sections. Building on prior work in technology design and adoption \cite{davis_technology_2024, scherer_technology_2019, xu_developing_2022, williams_social_1996}, we synthesize three higher-level influences -- Sociocultural, Institutional, and Technological -- that shaped the LISTED factors and ultimately determined how culturally aligned AI systems were built, adopted, and sustained. We then present twelve guidelines derived from our findings and situate them within this broader literature.

\begin{figure*}[t]
    \centering
    \includegraphics[width=\linewidth]{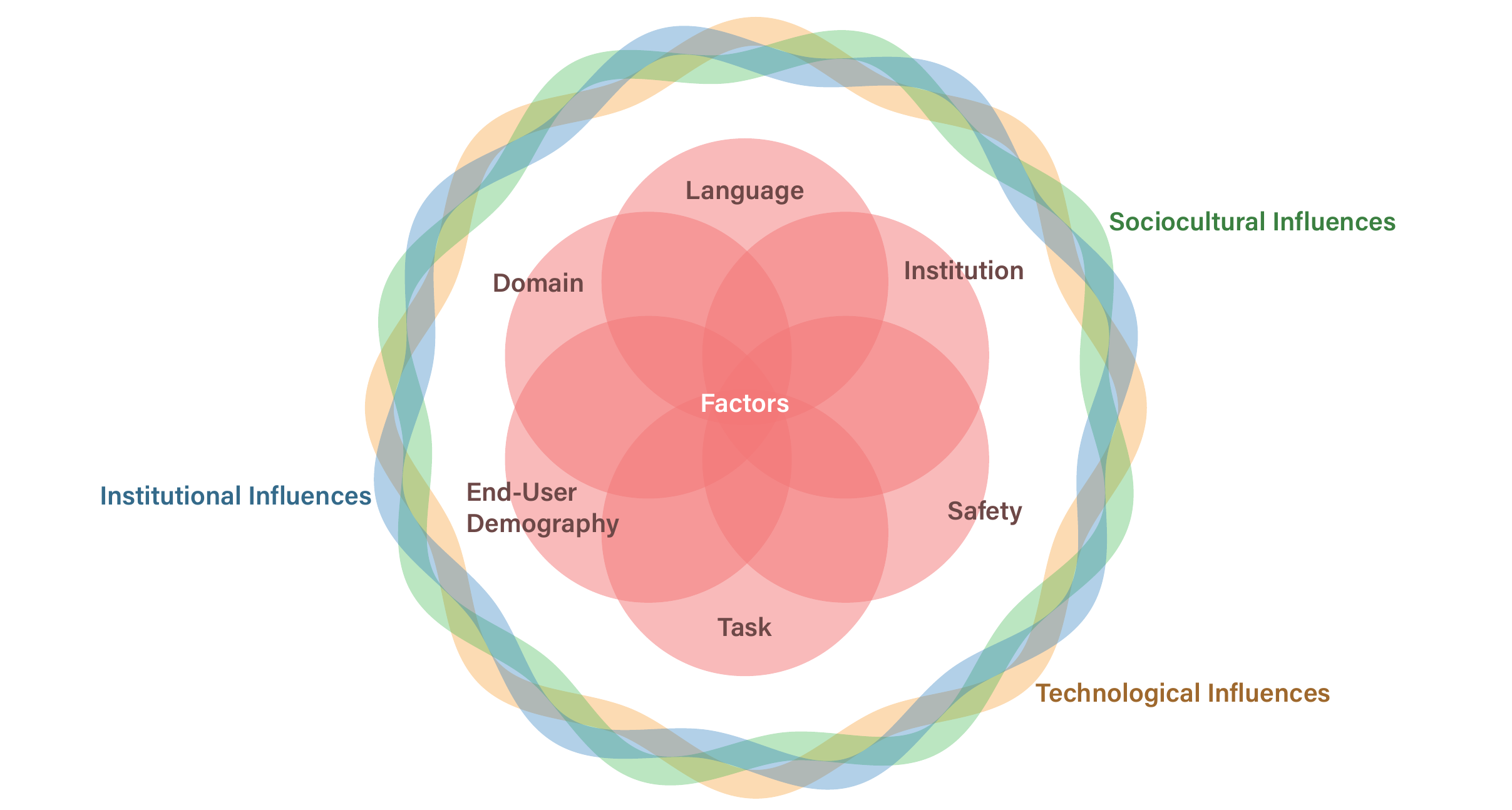}
    \caption{Factors and Influences shaping AI Systems for Social Good in non-Western Contexts} 
    \label{fig:factors_and_influences}
\end{figure*}



\subsection{Sociocultural Influences}
Sociocultural influences shaped how developers addressed Language, End-User Demography, Domain, and Safety factors, determining which communities were served and how trust was established. Prior work has long shown that sociocultural contexts shape technology design \cite{heimgartner_culturally-aware_2018, kim_cultural_2005, evers_cross-cultural_1999, heukelman_exploring_2009, medhi_text-free_2006}, but AI intensifies these dynamics. Traditional software has constrained inputs (forms, dropdowns, fixed workflows) and deterministic outputs, making cultural contextualization more bounded and predictable. In contrast, conversational AI systems accept open-ended, unstructured inputs (freeform text, speech, images) embedded with cultural and semantic symbols specific to local communities, and generate probabilistic outputs where the same query may yield different results. This unpredictability makes failures opaque and difficult to debug, and it can erode user trust in systems that behave inconsistently. 

\recommendation{Effective systems require AI developers and local domain experts to collaborate as equal partners throughout the full lifecycle of the application. }{Our findings show that culturally aligned AI systems emerged when AI developers and local domain experts contributed jointly across design, implementation, evaluation, and scaling. Local experts were not simply intermediaries for reaching users; they brought indispensable knowledge of community practices, language use, and on-the-ground workflows (\S\ref{language-variation-modality}). AI developers contributed complementary technical expertise that shaped model behavior, data pipelines, and system architecture (\S\ref{subsub:model-selection}). Across projects, the sociocultural complexity of deployments and the long-tailed input and output spaces shaped by community-specific language use required both forms of expertise to be present throughout the lifecycle. Nuances tied to Language and End-User Demography continued to surface as systems were used (\S\ref{subsub:granularity-challenge}), and teams relied on collaborative interpretation and decision-making to refine system behavior over time. Establishing structures that support shared ownership, ongoing dialogue, and continuous feedback enabled both groups to contribute fully and helped ensure that the system remains culturally grounded and responsive.}



\begin{table*}[t]
\added{
\centering

\label{tab:design-guidelines}

\begin{tabularx}{\linewidth}{
  >{\raggedright\arraybackslash}p{0.12\linewidth}  
  >{\raggedright\arraybackslash}p{0.02\linewidth}  
  >{\raggedright\arraybackslash}X                  
  >{\centering\arraybackslash}p{0.02\linewidth}    
  >{\centering\arraybackslash}p{0.02\linewidth}    
}
\toprule

\textbf{Influence} & \textbf{\#} & \textbf{Guideline} &  &  \\
\midrule

\multirow{4}{*}{\textbf{Sociocultural}}
& G1 & Effective systems require AI developers and local domain experts to collaborate as equal partners throughout the full lifecycle of the application
     & \aioff & \nw \\
\cmidrule(l){2-5}
& G2 & Culturally aware AI systems require dedicated attention to the linguistic and cultural gaps present in low-resource contexts
     & \ai & \nw \\
\cmidrule(l){2-5}
& G3 & Dominant languages often provide workable alternatives to unsupported local languages in contexts where communities are multilingual
     & \ai & \nw \\
\cmidrule(l){2-5}
& G4 & Community trust in AI is a key influence on system design in both mitigating doubt and preventing over-reliance
     & \ai & \nw \\
\midrule

\multirow{4}{*}{\textbf{Institutional}}
& G5 & Compliance with institutional mandates and workflows shapes whether AI systems achieve practical uptake
     & \aioff & \nwoff \\
\cmidrule(l){2-5}
& G6 & AI strengthens existing institutional capacities but cannot substitute for missing foundations
     & \ai & \nwoff \\
\cmidrule(l){2-5}
& G7 & Sustainable AI deployment relies on strong institutional capacity
     & \ai & \nw \\
\cmidrule(l){2-5}
& G8 & Institutional capacity shapes the depth of model adaptation, with in-context methods often preceding in-weight investments
     & \ai & \nwoff \\
\midrule

\multirow{4}{*}{\textbf{Technological}}
& G9 & Shifting model capabilities call for flexible system designs that can accommodate change without friction
     & \ai & \nwoff \\
\cmidrule(l){2-5}
& G10 & Small-scale deployments limit risk early by exposing failures that only emerge in real use
     & \ai & \nwoff \\
\cmidrule(l){2-5}
& G11 & Effective AI evaluation accounts for context that conventional benchmarks overlook
     & \aioff & \nwoff \\
\cmidrule(l){2-5}
& G12 & Effective Safety emerges from combining lightweight models, human-in-the-loop workflows, and frontier models rather than relying on technology alone
     & \ai & \nw \\
\bottomrule
\end{tabularx}

\vspace{0.5em}
\begin{minipage}{0.9\linewidth}
\footnotesize
\centering
\ai\; AI \quad
\nw\; Non-Western

\vspace{1em}

\textit{While the above guidelines are broadly applicable to software projects across contexts, the icons highlight those that carry heightened relevance for AI projects and for work in non-Western contexts.}
\end{minipage}
\vspace{1em}

\caption{Design Guidelines for Building Culturally Aligned AI Systems in Non-Western Contexts}
}
\end{table*}


\recommendation{Culturally aware AI systems require dedicated attention to the linguistic and cultural gaps present in low-resource contexts. }{Our findings show that teams working in non-Western settings often needed to invest substantial effort simply to achieve basic language functionality. In high-resource languages such as English, models generally perform well out of the box \cite{li_language_2025}, so most work focuses on refining domain-specific behavior. In contrast, many communities in our cases relied on low-resource languages or dialects where baseline model performance was poor, and the underlying NLP inequities were especially visible \cite{ranathunga_languages_2022, joshi_state_2021, bhagat_richer_2024, mirza_global-liar_2024}. Addressing these foundational gaps required curating dialect-sensitive vocabularies (\S\ref{subsub:granularity-challenge}), developing human review processes to check meaning and translation quality (\S\ref{human-in-the-loop}), and instruction-tuning models to better support under-served linguistic and demographic groups (\S\ref{language-variation-modality}). These activities demanded ongoing human and financial resources rather than one-time fixes, and teams that accounted for these investments early on were better able to build systems that remained usable and culturally grounded across diverse contexts.}

\recommendation{Dominant languages often provide workable alternatives to unsupported local languages in contexts where communities are multilingual. }{Our findings show that even with substantial effort, fully supporting some marginalized languages was not feasible because they were unsupported by existing AI infrastructures \cite{joshi_state_2021}. In these situations, teams relied on the multilinguality that was common in many non-Western contexts, where communities often spoke both a local language and a more dominant regional language \cite{karusala_only_2018, da_rosa_designing_2022, da_rosa_investigating_2024, choi_toward_2023}. Rather than attempting full modeling of severely under-resourced languages, teams adapted the dominant language used by the same population. For example, \farmerchat worked with Hindi but introduced Bhojpuri-relevant glossaries and phrasing adjustments to create more usable interactions for Bhojpuri-speaking farmers (\S\ref{language-variation-modality}). These cases show that when direct support for a marginalized language is out of reach, building thoughtful adaptations on top of a closely related or widely spoken alternative can improve accessibility while remaining grounded in the community’s linguistic reality.}



\recommendation{Community trust in AI is a key influence on system design in both mitigating doubt and preventing over-reliance. }{Our findings revealed that the trust in AI varied widely across communities and domains, echoing prior work that identifies capability, usability, and social influence as antecedents of trust \cite{dang_unveiling_2025, bach_systematic_2024, barnes_ai_2024, liu_is_2023}. Across our cases, trust emerged as a sociocultural process shaping how systems were structured and deployed. In \cataractbot, patients expressed confidence in responses because they were reviewed by doctors, aligning system workflows with cultural expectations of medical authority (\S\ref{human-in-the-loop}). By contrast, \suvas included disclaimers noting that AI-generated legal translations had no legal standing (\S\ref{subsub:mesaures-during-processing}) to prevent users from placing unwarranted trust in the outputs. These cases illustrate how system design was shaped by community perceptions of authority, expertise, and risk, requiring attention to both skepticism and over-reliance in context.}



\subsection{Institutional Influences}
Institutional influences shaped how AI systems were adapted, governed, and sustained, often defining the limits of what was possible beyond technical ambition. Prior work shows that AI adoption depends heavily on the institutional environments in which systems operate \cite{sheikh_contextualization_2023, patnaik_exploring_2024}, and our cases reinforce this, particularly in high-stakes domains where institutions determined how responsibilities were shared between humans and technologies. Consistent with research on AI use in business \cite{kurup_factors_2022, horani_critical_2025, erdmann_influence_2025}, top leadership support from governments and NGOs was pivotal for adoption. Institutions, however, varied widely in technical capacity, staffing, and resources, shaping how deeply models could be adapted, how much human expertise could be sustained, and whether systems could scale over time. These dynamics highlight the need to design for institutional fit: choosing adaptation strategies that match capacity, budgeting for ongoing human labor, and planning for long-term maintenance rather than one-time builds.

\recommendation{Compliance with institutional mandates and workflows shapes whether AI systems achieve practical uptake. }{Our cases echoed prior research that highlights the importance of aligning new technologies with institutional mandates and workflows \cite{davis_technology_2024, rogers_diffusion_2008}. Despite the novelty of AI and the clear productivity gains it offered, institutional alignment remained non-negotiable for successful adoption. In \shiksha, teachers’ initial enthusiasm only translated into use once the system’s outputs matched state-prescribed lesson plan formats and curricular expectations, and integrated with the lesson planning workflows of the teachers (\S\ref{subsub:institutional-requirement}). A similar pattern appeared in health deployments. In \farmerchat, the advice provided by the chatbots had to be aligned with the government advisories.  These cases show that improvements in service quality were not enough on their own; AI tools gained traction only when they fit smoothly into established institutional routines rather than creating new ones.}

\recommendation{AI strengthens existing institutional capacities but cannot substitute for missing foundations. }{Across our cases, AI systems performed well only when they were embedded within institutions that already had basic structures, workflows, and staffing to support their use. The technology enhanced existing processes but could not compensate for missing or overstretched institutional capacity echoing the argument by \citet{toyama_geek_2015}. In \cataractbot, the system worked effectively because doctors were already supporting patients through WhatsApp groups, and the AI simply streamlined an established workflow (\S\ref{human-in-the-loop}). By contrast, \ashabot struggled because the chatbot introduced additional demands that supervisors did not have the capacity to meet. Supervisors were expected to respond to questions the bot could not answer, but their existing workloads and occasional knowledge gaps made this impractical, leading to delayed or absent support. These cases show that AI systems can improve efficiency and reach, but they depend on baseline institutional capacity and cannot act as substitutes for missing infrastructure or human expertise.}

\recommendation{Sustainable AI deployment relies on strong institutional capacity. }{Our findings showed that sustaining AI systems in practice required continuous rather than one-time commitments, unlike many traditional software deployments that demand relatively modest upkeep. Language support needed ongoing updates as user demographics expanded and new gaps surfaced (\S\ref{subsub:model-selection}). Domain knowledge bases required periodic expert curation across applications (\S\ref{reliance-on-curated-knowledge}), and Safety oversight, including the validation workflows in \suvas, depended on sustained staffing (\S\ref{subsub:mesaures-during-processing}). Model-related costs also increased as user bases grew, since higher traffic raised inference expenses, infrastructure requirements, and monitoring needs to maintain quality and safety. While prior work has emphasized institutional support and compatibility as enablers of AI adoption \cite{kurup_factors_2022, horani_critical_2025, erdmann_influence_2025}, our cases show that long-term financial and human capacity strongly shaped whether systems remained viable once deployed and whether they could scale responsibly over time.}

\recommendation{Institutional capacity shapes the depth of model adaptation, with in-context methods often preceding in-weight investments. }{Decisions about whether to rely on in-context methods or pursue deeper in-weight modifications were shaped primarily by what institutions could realistically sustain. In-weight approaches such as fine-tuning or training new models offered performance gains but required maintaining specialized infrastructure, updating models as technology evolved, and providing ongoing oversight — demands that were especially challenging in low-resource language settings with limited public data \cite{mcgiff_overcoming_2025, kshetri_linguistic_2024}. In practice, institutions gravitated toward in-context approaches like prompt engineering and glossaries because they were faster to implement, easier to adjust as conditions changed, and more feasible given existing capacity. Projects such as \padhai and \prompts adopted in-weight modifications only when in-context approaches no longer met their needs (\S\ref{language-variation-modality}). These patterns show that in-context adaptations often provided an effective, sustainable middle ground for progressing without taking on burdens institutions could not maintain.}

\subsection{Technological Influences}
Technological influences shaped how developers addressed all LISTED factors, determining not only what teams aimed to build but also what was feasible at any moment. Because AI capabilities evolve rapidly and often through experimental progress \cite{tang_pace_2020}, developers worked within shifting constraints of model performance, evaluation practices, and resource availability. While these technologies offer new opportunities for expanding access to information and services, scholars caution that they can also reinforce global inequities and cultural harms \cite{bengio_managing_2024}. Our cases show how both the benefits and risks of AI become tangible in non-Western contexts, where uneven model performance across languages, cultures, and tasks \cite{agarwal_ai_2025, hershcovich_challenges_2022, qin_survey_2025} directly shapes the design of culturally aligned systems for social good. These technological dynamics required teams to adapt pragmatically through strategies such as anticipating rapid model improvements, developing localized evaluation methods, deploying early to surface real-world gaps, and maintaining strong human fallbacks when automation fell short.

\recommendation{Shifting model capabilities call for flexible system designs that can accommodate change without friction. }{Across our cases, developers operated in a technological landscape evolving faster than traditional development cycles. Model capabilities improved rapidly, making heavy investment in engineering fixes for current limitations inefficient when those limitations were likely to be resolved by broader advances. Teams therefore favored modular designs that allowed new models or components to be integrated without rebuilding entire pipelines. Developers adapted pragmatically to what the technology could support at each moment. In \shiksha, teams combined multiple models to balance strengths in content generation and translation (\S\ref{subsub:model-selection}). In \suvas, early deployments relied on NMT because LLMs were not yet widely available. In \prompts, commercial frontier models eventually outperformed the continually pretrained in-house Swahili model. Together, these cases show that system design must anticipate rapid capability shifts and remain flexible enough to incorporate new models with minimal redevelopment.}

\recommendation{Small-scale deployments limit risk early by exposing failures that only emerge in real use. }{Early limited deployments played a critical role in managing technological risk because many AI failures surfaced only when systems interacted with the diversity and unpredictability of real users. These failures stemmed not from coding errors but from the probabilistic and context-dependent nature of model behavior, which made pre-deployment testing insufficient. Small rollouts created a safer environment to observe how models handled long-tailed language inputs, community-specific references, and domain nuances absent from development settings. In \farmerchat, for instance, early deployments revealed that the STT model mis-transcribed similar-sounding words, leading to incorrect agricultural advice — problems detectable only when exposed to real accents, pronunciations, and vocabulary (\S\ref{subsub:model-selection}). Across cases, small-scale deployments served as crucial risk-buffering phases, surfacing hidden vulnerabilities before systems reached larger populations.}

\recommendation{Effective AI evaluation accounts for context that conventional benchmarks overlook. }{Dominant AI evaluation paradigms often reflected technological assumptions that conflicted with ground realities. Standard benchmarks used for model selection \cite{hardy_more_2025} were frequently misaligned with local contexts because of the assumptions embedded in their datasets \cite{reuel_betterbench_2024, eriksson_can_2025, mcintosh_inadequacies_2025}. Developers treated these benchmarks as starting points but quickly encountered their limits, echoing critiques by \citet{liao_rethinking_2025}. Projects therefore created localized evaluation methods. \farmerchat developed gold-standard transcripts to assess agricultural terminology (\S\ref{subsub:model-selection}), and user feedback loops became central to evaluating Safety and cultural relevance across diverse End-User Demography (\S\ref{subsub:granularity-challenge}). These cases show that evaluation practices had to evolve toward participatory, context-sensitive approaches that captured nuances generic benchmarks could not represent.}

\recommendation{Effective Safety emerges from combining lightweight models, human-in-the-loop workflows, and frontier models rather than relying on technology alone. }{Across our cases, Safety depends on multiple layers working together rather than on any single technological component. Lightweight models, state-of-the-art LLMs, and human expertise each address different aspects of risk, and systems remain dependable only when these elements are integrated. In \prompts, which handles nearly 12,000 questions per day, a lightweight classifier screens each query for sensitive or high-risk content before it reaches the LLM, enabling the AI system to safely handle around 80\% of incoming messages. Queries involving critical or high-risk issues are then routed to a dedicated human help desk whose judgments cannot be replaced by automated methods alone. These workflows function effectively because they draw on existing institutional processes, with AI augmenting rather than substituting human capacity. Across deployments, this pattern shows that robust Safety in complex, multilingual, and high-stakes settings arises from combining lightweight triage models, frontier models for general reasoning, and human oversight for critical and high-risk cases.}



Together, these 12 guidelines underscore that culturally aware AI systems cannot be achieved through technical fixes alone. Effective system design should address the LISTED factors that are shaped by Sociocultural, Institutional, Technological influences to ensure that systems remain usable and trustworthy in high-stakes non-Western contexts.
}

\section{Conclusion}\label{sec:conclusion}
We conducted semi-structured interviews with AI developers and domain experts who have built and deployed AI for social good in education, healthcare, agriculture, and law across non-Western contexts. Our findings reveal six critical factors that consistently influence how these systems are built and deployed: \added{Language, Institution, Safety, Task, End-User Demography, and Domain.} We document how adaptation strategies varied from lightweight, in-context modifications to more substantial in-weight changes. \added{These technical choices were shaped not only by computational factors but also by Sociocultural, Institutional, and Technological influences.} Most significantly, our findings underscore the extensive human labor required for successful deployment. AI developers and domain experts engaged in intensive collaboration throughout the process such as curating datasets, implementing human-in-the-loop systems, and designing safety mechanisms tailored to local contexts. We identify where this human involvement proved most essential, revealing how labor-intensive practices became fundamental to achieving relevance, safety, and trust in real-world applications. Our work concludes with \added{12 actionable guidelines for practitioners} and policymakers aimed at creating AI for social good systems in non-Western contexts that are contextually appropriate, equitable, and responsive to diverse community needs.

\begin{acks}
We would like to thank our participants in the study for their time and thoughtful inputs. We thank Global Cornell and Cornell Global AI initiative for supporting this work. We also thank anonymous reviewers for their constructive and respectful feedback on our work.
\end{acks}

\bibliographystyle{ACM-Reference-Format}
\bibliography{references}

\pagebreak
\appendix
\section{AI Adaptation Techniques}\label{appendix:adaptation-techniques}

\noindent
\incontext{Prompt Engineering} \\
Prompt Engineering is the crafting and iteratively refining of natural language prompts—sometimes with examples—to guide generative models toward more accurate, relevant, and controlled outputs \cite{sahoo_systematic_2025}.
\bigskip

\noindent
\rag \\
Retrieval-Augmented Generation (RAG) enhances LLM responses by retrieving and grounding them in external knowledge sources (e.g., documents or databases), reducing hallucination and improving factual reliability \cite{NEURIPS2020_6b493230}.
\bigskip

\noindent
\glossaryadapt \\
Glossary Adaptation incorporates domain-specific vocabulary or local terminology—via glossaries—deterministically into the inputs and outputs of the AI system to ensure appropriate and culturally relevant expressions \cite{ghazvininejad_dictionary-based_2023}.
\bigskip

\noindent
\humanintheloop \\
Human-in-the-Loop involves routing model outputs—especially high-impact or uncertain ones—to human experts for review, correction, or confirmation before delivering them to end users \cite{mosqueira-rey_human---loop_2023}.
\bigskip

\noindent
\modeltraining \\
Model Training builds models from scratch using domain-relevant data—e.g., training ASR on children’s speech—to create capabilities tailored to specific user needs \cite{fan_towards_2022}.
\bigskip

\noindent
\continualpretrain \\
Continual Pre-training continuously updates a pre-trained model with new domain- or language-specific data to enhance its performance in evolving contexts \cite{liu_culturally_2025}.
\bigskip

\noindent
\finetuning \\
Fine Tuning further trains a pre-trained model on labeled, domain-specific datasets (e.g., legal text) to better align its behavior with specialized tasks and terminology \cite{liu_culturally_2025}.
\bigskip

\noindent
\instructiontuning \\
Instruction Tuning adjusts a model using instruction–response pairs so it better understands and follows human-written directives across tasks and styles \cite{liu_culturally_2025}.
\bigskip

\section{AI Developer Interview Protocol}\label{appendix:ai-dev-interview-protocol}

\section*{Introduction (5 mins)}

Thank you for taking the time to participate in this interview. We are conducting research on how AI systems are being built for social good applications. We would like to learn more about your experience developing AI solutions that address social challenges. This interview will be approximately 60--75 minutes long. 

\medskip
\noindent
Do I have your consent for this interview and permission to record this conversation?

\section*{Demographics (5 mins)}
\begin{itemize}
    \item What is your current role and organization?
    \item Age, Gender, Education
    \item How many years of experience do you have in AI development?
    \item Can you briefly describe the AI project(s) for social good that you've worked on? What specific social problem does each of your AI systems address?
\end{itemize}

\section*{Problem Context \& Technology Selection (15 mins)}
\begin{itemize}
    \item What led you to choose AI as the solution approach for this particular problem?
    \item How was this problem being managed or solved before your AI intervention?
    \item What unique advantages does AI provide compared to other technological solutions you considered?
    \item What are the disadvantages of using AI for solving this problem?
    \item What are the non-negotiable design principles or constraints that guide your technical decisions?
    \item How does designing AI solutions for social good differ from developing commercial AI applications?
\end{itemize}

\section*{Contextualization \& Cultural Adaptation (15 mins)}
\begin{itemize}
    \item What did you do to make your intervention culturally and locally relevant to communities on the ground?
    \item How do you technically implement this contextualization? What parts of your system require cultural adaptation?
    \item Who is involved in the contextualization process, and how do you incorporate their input into technical decisions?
    \item Can you provide specific examples of how cultural context influenced your technical architecture?
    \item How do you measure whether your contextualization efforts are effective?
    \item How do you engage with target communities throughout the technical development process?
    \item What methods have you found most effective for incorporating community feedback into your technical work?
    \item How is the diversity of the population captured in the contextualization process?
    \item What technical challenges do you encounter when working with social impact practitioners?
    \item How is your solution technically delivered to target communities? What infrastructure or accessibility considerations shaped your approach?
\end{itemize}

\section*{AI Harms \& Bias Mitigation (15 mins)}
\begin{itemize}
    \item What specific limitations have you identified in the AI models you're using?
    \item What biases have you identified in the AI models you're using?
    \item What potential harms have you identified in the AI models you're using?
    \item What concrete mitigation strategies have you implemented to address these issues?
    \item How do you test the effectiveness of your bias mitigation approaches?
    \item How do you test the effectiveness of your harm mitigation approaches?
\end{itemize}

\section*{Development Process (15 mins)}

\begin{itemize}
    \item Can you walk me through your project development lifecycle from start to finish?
    \item Who is involved at each stage of development, and what are their roles?
    \item How do you define and measure the success of your technical intervention?
    \item What are the biggest gaps that you see in the system that is limiting the success of the interventions?
\end{itemize}

\section*{Wrap Up}

\begin{itemize}
    \item What would you do differently if you were starting this project over?
\end{itemize}

\section{Domain Expert Interview Protocol}\label{appendix:domain-expert-interview-protocol}

\section*{Introduction (5 mins)}

Thank you for taking the time to participate in this interview. We are conducting research on how AI systems are being built for social good applications. We would like to learn more about your experience implementing AI solutions that address social challenges. This interview will be approximately 60--75 minutes long. 

\medskip
\noindent
Do I have your consent for this interview and permission to record this conversation?

\section*{Demographics (5 mins)}

\begin{itemize}
    \item What is your current role and organization?
    \item Age, Gender, Education
    \item How many years of experience do you have in social impact work?
    \item Can you briefly describe the AI project for social good that you've been involved with?
\end{itemize}

\section*{Problem Understanding \& Solution Approach (15 mins)}

\begin{itemize}
    \item What specific social problem does your AI intervention address?
    \item How was this problem being managed by your organization before implementing AI?
    \item What factors led your organization to decide to use AI for this problem?
    \item What unique value does AI add that other approaches couldn't provide?
    \item What are the disadvantages of using AI for solving this problem?
    \item How do the advantages weigh over the disadvantages?
    \item What are the limitations of the current AI models? What more capabilities are needed to solve the problem more aptly?
    \item How does this AI intervention fit into your organization's broader work?
    \item What are the short-term and long-term effects of using the tool?
\end{itemize}

\section*{Contextualization \& User Relevance (15 mins)}

\begin{itemize}
    \item How did you ensure that the AI intervention is culturally appropriate to local end-user communities?
    \item How do you ensure the technology is fully relevant for your end users?
    \item What do the users primarily value?
    \item How do you ensure that users' values are reflected in the AI technologies?
    \item Who are the main stakeholders in the intervention?
    \item What do you do when values conflict between different stakeholders?
    \item What processes do you have in place to assess cultural relevance and appropriateness?
    \item How is the diversity of the population captured in the contextualization process?
    \item What support mechanisms do you provide to help users better utilize the AI tool?
    \item What training is provided to support better adoption?
\end{itemize}

\section*{Partnership \& Collaboration (15 mins)}

\begin{itemize}
    \item How do you collaborate with the technology team throughout development and implementation?
    \item What does effective collaboration between social impact practitioners and technologists look like?
    \item How do you engage with end users throughout the implementation process?
    \item What challenges do you face when coordinating between technologists, end users, your organization, and the government?
    \item What are the main tensions you've encountered in these partnerships, and how do you navigate them?
    \item What opportunities have emerged from working across these different partnerships?
\end{itemize}

\section*{Implementation \& Adoption (15 mins)}

\begin{itemize}
    \item What are the main barriers to adoption you encounter with your target users?
    \item How do you systematically address these adoption challenges?
    \item Do they find the tool useful? And in what way?
    \item How do you measure the impact of your AI intervention?
    \item What potential harms do you see from using AI to solve this particular problem?
    \item How do you monitor and mitigate these risks?
\end{itemize}

\section*{Wrap-up (5 mins)}

\begin{itemize}
    \item What would you do differently if you were starting this project over?
    \item What advice would you give to other organizations attempting similar AI for social good work?
    \item Are there any other aspects of your experience that we haven't discussed that you think are important?
    \item Do you have suggestions for other practitioners we should speak with for this research?
\end{itemize}

\end{document}